\documentclass[10pt,twocolumn]{IEEEtran}
\usepackage[T1]{fontenc}
\usepackage{amsmath,amssymb}
\usepackage{graphicx}
\usepackage{epsfig}
\usepackage{cite}
\usepackage{setspace}
\usepackage{amsthm}
\usepackage{color}
\usepackage{enumerate}

\newcommand{\mB}{\mathcal B}
\newcommand{\mA}{\mathcal A}
\newcommand{\mR}{\mathcal R}
\newcommand{\mF}{\mathcal F}
\newcommand{\tQ}{\tilde Q}
\newcommand{\hQ}{\hat Q}
\newcommand{\tL}{\tilde L}
\newcommand{\hL}{\hat L}
\newcommand{\tD}{\tilde D}
\newcommand{\hD}{\hat D}

\newcommand{\mT}{\mathcal T}

\newtheorem{mydef}{Definition}
\newtheorem{lemma}{Lemma}

\newtheorem{prop}{Proposition}

\title{A Game Theoretic Model for the Gaussian Broadcast Channel}
\author{Srinivas Yerramalli, Rahul Jain and Urbashi Mitra \\
       University of Southern California \\
      \{srinivas.yerramalli,rahul.jain,ubli\}@usc.edu
      \thanks{This research has been funded in part by the following grants and organizations: NSF CCF-0917343, NSF IIS 0917410, CAREER CNS-0954116, ONR N00014-09-1-0700, NSF CNS 0832186, NSF CCF-1117896 and NSF CNS-082175}}
      
\begin{document}
\maketitle
\pagestyle{empty}
\thispagestyle{empty}

\begin{abstract}
THIS PAPER IS ELIGIBLE FOR THE STUDENT PAPER AWARD. The behavior of \textit{rational} and \textit{selfish} players (receivers) over a multiple-input multiple-output Gaussian broadcast channel is investigated using the framework of non-cooperative game theory. In contrast to the game-theoretic model of the Gaussian multiple access channel where the set of feasible actions for each player is independent of other players' actions, the strategies of the players in the \textit{broadcast channel} are mutually coupled, usually by a \textit{sum power} or \textit{joint covariance} constraint, and hence cannot be treated using traditional Nash equilibrium solution concepts. To characterize the strategic behavior of receivers connected to a single transmitter, this paper models the broadcast channel as a generalized Nash equilibrium problem with coupled constraints. The concept of normalized equilibrium (NoE) is used to characterize the equilibrium points and the existence and uniqueness of the NoE are proven for key scenarios.

\end{abstract}

\section{Introduction}
\label{sec:intro}
We consider a multiple-input multiple-output (MIMO) Gaussian \textbf{broadcast channel} (GBC) in which a single transmitter is assigned to send independent messages to several receivers subject to a sum power constraint or a joint covariance constraint. \textit{Our goal is to introduce a game-theoretic model for signaling over the GBC to model strategic non-cooperative behavior among receivers, who are the players in the game}. Such a model allows us to predict the operating point of a broadcast channel, (\textit{e.g.,} finding the equilibrium precoding matrices), and enables the development of distributed algorithms to achieve this operating point.

In a typical non-cooperative game, the choice of an action by a player affects the utility obtained by \textit{every} player, but does not change the set of available actions for other players. Several game theoretic models for signaling over wireless channels adopt this framework for characterizing the interaction between rational or selfish nodes (see \cite{Scutari10,survey11} for a detailed survey). For example, the Gaussian multiple access channel (MAC) has been studied extensively under this framework \cite{La04,Lai08,Belmega10,Srinivas11ISIT}. In practice, we encounter several scenarios in which the choice of actions of one player may modify the set of feasible choices for every other player \cite{Rosen65,Facchinei07}. For example, in \cite{Pang08intfchannel}, the problem of maintaining a minimum rate over parallel Gaussian interference channels subject to a sum power constraint for each user is considered. The choice of power allocation of a player for a given interference channel is then also influenced by the choice of power allocation by other players on other interference channels to maintain an overall desired rate and hence is a game with coupled constraints. For the Gaussian broadcast channel, the choice of precoding covariance matrix for one of the receivers restricts the choice of precoding matrices for all other receivers, due to the joint power or covariance constraints. Interaction of players at the level of feasible sets makes the problem much harder than standard non-cooperative games. The problem of determining the equilibrium points of games with coupled constraints is called the generalized Nash equilibrium problem (GNEP) \cite{Scutari10,Pang08intfchannel,Rosen65} and the points themselves are called generalized Nash equilibria (GNE).

From a game theoretic perspective, the broadcast channel has received very little attention as compared to other channels such as the MAC. A discrete memoryless broadcast channel with 2 users and a resource manager was considered in \cite{Su08} and impact of the information available to the resource manager in modifying utility of each user is studied.

The main contributions of this paper are the following: (1) We model the interaction between selfish receivers getting data over a GBC, with the transmitter employing dirty paper coding or linear precoding, as a GNEP. The transmitter computes the precoding matrix for the message intended for each receiver and the precoding matrices are constrained by a common joint constraint. (2) We first show that there exists at least one GNE for the broadcast channel. In fact, it is well know that under mild conditions there may exist infinitely many GNEs for a given GNEP \cite{Rosen65,Facchinei07}. Games in which each player has the same coupled constraints, such as in the broadcast channel game, belong to a special class whose GNEs can be characterized with weight vectors. The GNEs of this special class, first defined by Rosen \cite{Rosen65}, are called normalized equilibrium (NoE) points. (3) Using Rosen's methodology, we show that for every feasible weight vector there exists at least one NoE. (4) For the special case of the aligned and degraded broadcast channel \cite{Weingarten06}, we first derive a sufficient condition for the uniqueness of NoEs as was done in \cite{Belmega10} for the multi-antenna Gaussian MAC (see also \cite{Rosen65}) and then (5) determine the set of weight vectors for which the NoE is unique.

The rest of the paper is organized as follows. Section \ref{sec:signal_model} briefly introduces the model for the broadcast channel and presents several game theoretic definitions and preliminaries useful for the rest of the paper. Section \ref{sec:game_model} discuss the existence and uniqueness of GNEs and Section \ref{sec:conclusions} concludes this paper. 

\section{Signal Model and Preliminaries}
\label{sec:signal_model}
In this section, we describe the signal model and state several game theoretic concepts used throughout this paper.

\subsection{Broadcast channel model}
Consider a general multiuser multi-input multi-output (MIMO) system with one transmitter and $K$ receivers. We first define several classes of broadcast channels which are simplified models of the GBC. We use the terminology from \cite{Weingarten06} herein. The transmitted signal is denoted by $\underline{x}$, a vector of length $t$, where $t$ is the number of antennas at the transmitter (TX). This TX signal is the sum of independent  $\underline{x_i}$, each drawn from a Gaussian codebook and intended for the $i^{th}$ receiver (RX).
\begin{equation}
\underline{x} = \sum_{i=1}^{K} \underline{x}_i, ~ \underline{x}_i \sim \mathcal{N}(0,Q_i) .
\end{equation}
The signal at the $i^{th}$ RX can be expressed as
\begin{equation}
\underline{y}_i = H_i \underline{x} + \underline{n}_i, ~ \underline{n} \sim \mathcal{N}(0, \sum_{i=1}^{K} N_i),
\end{equation}
where $H_i$ is the $ r_i \times t $ channel gain matrix from the TX to the $i^{th}$ RX and $r_i$ is the number of antennas at the $i^{th}$ RX. Without loss of generality, we assume that transmitter signaling is constrained by a covariance matrix $ {\bf S} \succeq 0$ such that
\begin{equation}
E[ \underline{x} \underline{x}^{T} ] = \sum_{i=1}^{K} Q_i  \preceq {\bf S}.
\end{equation}
The scenario with a sum power constraint, \textit{i.e.,} can be similarly modeled as
\begin{equation}
\mbox{Tr} \left( E[ \underline{x} \underline{x}^{T} ] \right) \leq P_{tot},
\end{equation}
where $P_{tot}$ is the maximum transmit sum power for all the antennas. For simplicity of illustration, we only consider the covariance matrix constraint in this paper; all the results derived are also valid for sum power constraint as well. We now define a special class of broadcast channels which are simplified versions of the general GBC \cite{Weingarten06}.

\subsubsection{AMBC} A MIMO BC is called \textit{aligned} if the number of transmit antennas is equal to the number of antennas at each of the receivers ($t = r_1 = r_2 = \ldots = r_K$) and the channel gain matrices are all identity matrices ($H_i = I_{t \times t} $).

\subsubsection{ADBC} A MIMO BC is called \textit{aligned} and \textit{degraded} if the BC is aligned and the covariances of the Gaussian noise at the receiver are ordered such that $ 0 \prec N_1 \preceq N_2 \preceq ... \preceq N_K $, where $ A \preceq B $ implies that $B-A$ is a positive semi-definite matrix. 

\subsection{Generalized Nash Equilibrium Problems}
\subsubsection{Definition}
Formally, a GNEP consists of $K$ players with each player controlling the variables $Q_k$ ($Q_k$ are positive semi-definite covariance matrices in our problem). Each player has an objective function, $v_k$, which depends on $Q_k$ as well as the controlling variables of all other players denoted by $Q_{-k}$. We denote the utility function by $v_k(Q_k,Q_{-k})$ or $v_k(Q), ~ Q = (Q_1,Q_2,...,Q_K)$ to emphasize the dependence on the controlling variables and will be called the \textit{utility} function of the $k^{th}$ player for the  rest of this paper. Given the strategies $ Q_{-k} $ picked by all other players, the set of feasible actions of the $k^{th}$ player determined by the joint constraints is given by $ \mA(Q_{-k}) $ and this is the \textit{feasible set} or the \textit{strategy space} of the $k^{th}$ player. We emphasize that the set $A_k(Q_{-k})$ is a function of the strategies of the other players. Each player, given the strategies of all other players, picks a strategy that solves the maximization problem
\begin{equation}
\max_{Q_k} v_k(Q_k,Q_{-k}) ~ \textrm{subject to} ~ Q_k \in \mA_k(Q_{-k}).
\label{e:GNEP}
\end{equation}
Let $ \mB_k (Q_{-k}) $ denote the set of all the solutions for this maximization problem. The GNEP is the problem of finding $Q^{*}_{k}$ such that
\begin{equation}
Q^{*}_k \in \mB (Q^{*}_{-k}) ~ \textrm{for all} ~ k = 1,2,...,K.
\end{equation}
Define $ \mB(Q) = \times_{i=1}^{K} \mB(Q_{-i}) $. In other words, the GNEP is to find a fixed point $Q^{*}$ such that $ Q^{*} \in \mB(Q^{*}) $ and $Q^{*}$ solves the maximization in (\ref{e:GNEP}). Such a point is called a \textit{generalized Nash equilibrium} (GNE) or a solution to the GNEP. A point $Q^{*} = (Q_1^{*},...,Q^{*}_{K})$ is therefore an equilibrium if no player can increase his objective function by unilaterally changing $Q_k$ to any other feasible point.

\subsubsection{Discussion} A GNEP usually has multiple or even infinitely many solutions \cite{Rosen65}. A special class of GNEPs, called GNEP with shared common constraints was defined by Rosen \cite{Rosen65}. This class is characterized by dependent constraints that are common to all the players and our paper herein considers a problem belonging to this class. Rosen proposed a solution concept called the normalized equilibrium (NoE) to characterize GNEP belonging to this class. The NoE is a GNE for which the Lagrange multipliers (shadow prices) associated with the shared constraints are equal among all players up to constant factors, and its uniqueness is guaranteed under appropriate conditions \cite{Rosen65}.

GNEs are not self-enforceable like a Nash equilibrium as it is not feasible to assume that each player picks his strategy independently and the selected strategies satisfy the coupled constraints. However, GNEs have significant explanatory power and capture the characteristics of several real world problems as will be demonstrated later.

\subsection{Concave games, existence and uniqueness of normalized equilibrium points}
\begin{mydef}
A game is said to be \textbf{concave} if the set of allowed strategies of all the players is a convex set, the utility functions are concave in each players control variables and continuous in the control variables of all other players.
\end{mydef}

Let us denote by $\mF$ the set of jointly feasible strategies of the players, and by $\mF_i$ the projection of $\mF$ on the space from which the control variables for the $i^{th}$ player come from. Then $ \mF \subseteq \mF_1 \times \mF_2 ... \times \mF_K $, with equality satisfied when the control variables do not have any joint constraints. Let us define the function
\begin{equation}
f(P,Q,\underline{r}) = \sum_{i=1}^{K} r_i v_i(Q_1,...,Q_{i-1},P_i,Q_{i+1},...,Q_{K}),
\end{equation}
for a fixed vector $\underline{r} = (r_1,r_2,...,r_K) \in \mathbb{R}^{K}$. The K-tuple  $Q^{*} = (Q^{*}_1,...,Q^{*}_K)$ is a normalized equilibrium point if $Q^{*}$ satisfies the equivalent fixed point condition
\begin{equation}
Q^{*} = \arg \max_{P} f(P,Q^{*},\underline{r}),
\label{e:GNEP_fixed_pt}
\end{equation}
where the maximization is carried out over the convex set $\mF$.

For concave games, the existence of a normalized equilibrium point is guaranteed by Theorem 3 in \cite{Rosen65} for all vectors $r$ in the positive orthant. Note that for any given value of $r$, there could be multiple normalized equilibrium points. We now state the conditions for the uniqueness of the NoEs.

\subsubsection{Uniqueness of normalized equilibrium points}
Let
\begin{equation}
\sigma(Q,\underline{r}) = \sum_{i=1}^{K} r_i v_i(Q_i,Q_{-i}), ~ r_i > 0,
\end{equation}
be a weighted sum of the utilities of each player, where $Q$ are the control variables for all the players and $r$ is a vector containing a set of weights.
\begin{mydef}
The function
\begin{equation}
g(Q,\underline{r}) = \left [ \begin{array}{c} r_1 \nabla_1 v_1(Q_1,Q_{-1}) \\  r_2 \nabla_2 v_2(Q_2,Q_{-2}) \\ \vdots \\  r_K \nabla_K v_K(Q_k,Q_{-k}) \end{array}
           \right ],
\end{equation}
where $\nabla_i$ is the derivative w.r.t the $i^{th}$ players'control variables is called the \textbf{pseudo-gradient} of $\sigma(Q,r)$.
\end{mydef}

\begin{mydef}
Vector valued strategies \cite{Rosen65}: The function $\sigma(\underline{x},\underline{r})$ is called \textbf{diagonally strictly concave} (DSC) in vector valued strategies
for $\underline{x} \in \mF$ and a fixed $ \underline{r} \in \mathbb{R}^{K}$ if for every $ \underline{x}^0, \underline{x}^1 \in \mF $, we have
\begin{equation}
(\underline{x}^{1} - \underline{x}^{0})^{T} g(\underline{x}^{0},\underline{r}) + (\underline{x}^{0} - \underline{x}^{1})^{T}
                        g(\underline{x}^{1},\underline{r}) > 0.
\end{equation}
\end{mydef}
From Theorem 4 in \cite{Rosen65}, we know that for vector valued strategies, if $\sigma(\underline{x},\underline{r})$ is DSC for every $r \in \mR$,
where $\mR$ is a convex subset of the positive orthant, then for each $\underline{r} \in \mR$ the NoE is unique.

\begin{mydef}
Matrix valued strategies \cite{Belmega10}: The function $\sigma(Q,\underline{r})$ is called DSC in matrix valued strategies
for $Q \in \mF$ and a fixed $ \underline{r} \in \mathbb{R}^{K}$ if for every $ Q^0, Q^1 \in \mF $, we have
\begin{equation}
\mbox{Tr} \left [ (Q^{1} - Q^{0})^{T} g(Q^{0},\underline{r}) + (Q^{0} - Q^{1})^{T} g(Q^{1},\underline{r}) \right ] > 0.
\end{equation}
\end{mydef}
We also show that if, for matrix valued strategies, the DSC condition is satisfied for every $\underline{r} \in \mR$, then the NoE is unique for that
$\underline{r} \in \mR $.

\section{The Broadcast Channel as a Generalized Nash Equilibrium Problem}
\label{sec:game_model}
\subsection{Model of the game}
As mentioned previously, the broadcast channel has a single TX sending data to several RXs over the wireless channel. The receivers are the players of this game and it can be assumed that a fictitious agent of each RX is located at the TX plays and the game on behalf of the RX. We assume Gaussian codebooks are used for communication and each RX's control variable is the covariance matrix $Q_i \succeq$ of the signal $x_i$ intended for that receiver ${RX}_i$. All these players are constrained by a joint covariance constraint given by  $ \sum_{i=1}^{K} Q_i \preceq {\bf S} $. Thus, it is clear that the dependent constraints are common to each player and hence this game belongs to the special class of GNEPs which can be characterized by NoE points.

We consider both linear precoding \cite{Guthy10} and dirty paper coding (DPC)\cite{Weingarten06} based encoding schemes at the TX. For the DPC, we also assume a fixed encoding order at the TX without time-sharing between orders. Each player obtains a rate (the game utility) based on the choice of actions of all other players and the TX. For a general GBC, if the data streams are linearly precoded at the TX with covariance matrices $Q_i$, the utility function (rate achievable) of the $k^{th}$ player can be written as \cite{Guthy10}:
\begin{equation}
v_k(Q_k,Q_{-k}) = \mbox{log} \left ( \frac{| N_i + H_i (\sum_{i=1}^{K} Q_i) H^{H}_i |}{| N_i + H_i (\sum_{i \neq k} Q_i) H^{H}_i  |} \right ).
\label{e:util_precoding}
\end{equation}
Similarly, if the data streams are coded using DPC with encoding order $(K,K-1,...,1)$, then the utility function of the $k^{th}$ player can be written as \cite{Weingarten06}:
\begin{equation}
v_k(Q_k,Q_{-k}) = \mbox{log} \left ( \frac{| N_i + H_i (\sum_{i=1}^{k} Q_i) H^{H}_i |}{| N_i + H_i (\sum_{i=1}^{k-1} Q_i) H^{H}_i  |} \right ).
\label{e:util_DPC}
\end{equation}
It is easy to see that the utilities of the $k^{th}$ player, $v_k(Q_k,Q_{-k})$ in (\ref{e:util_precoding}) and (\ref{e:util_precoding}), are concave in $Q_k$ and continuous in the control variables of all the other players.

\subsection{Existence of Normalized Equilibrium Points}

\begin{prop}
The broadcast channel game with linear precoding or DPC is a concave game and hence for each weight vector in the positive orthant $\underline{r} = (r_1,r_2,...,r_K) \in \mathbb{R}_{++}^{K}$ there exists at least one normalized equilibrium point.
\end{prop}
\begin{IEEEproof}
Each players' control variable in the broadcast channel game is the signaling covariance matrix $Q_i$. By definition $Q_i \succeq 0$. In addition, the sum power constraint or the joint covariance constraint ensure that the set of jointly feasible strategies is compact and convex. The $k^{th}$ players' utility is concave in $Q_k$ and is continuous in $Q_{-k}$ and hence the broadcast channel game is a concave game. From Theorem 3 of \cite{Rosen65}, we know that a concave game has at least one NoE for every weight vector $\underline{r} \in \mathbb{R}_{++}^{K}$.
\end{IEEEproof}

\subsection{Uniqueness of Normalized Equilibrium Points}
We now derive the condition for the uniqueness of the equilibrium points. We start by assuming that for a given $\underline{r} \in \mathbb{R}_{++}^{K}$ there exist multiple equilibrium points and then arrive at a contradiction which proves the uniqueness.

\begin{prop}
The sufficient condition for the uniqueness of the normalized Nash equilibrium for a given weight vector $\underline{r}$ is given as
\begin{equation}
\mbox{Tr} \left [ (\hQ - \tQ)^{T} g(\tQ,\underline{r}) + (\tQ_i - \hQ_i)g(\hQ,\underline{r}) \right ] > 0.
\end{equation}
\end{prop}
\begin{IEEEproof}
Let $ \tQ = \left ( {\tilde Q}_1,{\tilde Q}_2,...,{\tilde Q}_K \right) $ and $\hQ = \left ( {\hat Q}_1,{\hat Q}_2,...,{\hat Q}_K \right) $ be two K-tuples of covariance matrices which are normalized equilibria to the game characterized by the weight vector $r$. We know from (\ref{e:GNEP_fixed_pt}) that $f(\tQ,\tQ,\underline{r}) = \max_{P \in \mF} f(P,\tQ,\underline{r})$ and $f(\hQ,\hQ,\underline{r}) = \max_{P \in \mF} f(P,\hQ,\underline{r})$. Writing the Karush-Kuhn-Tucker (KKT) conditions \cite{Boyd} for the two equilibria yields:
\begin{enumerate}[(a)]
\item $\tQ_i,~ \hQ_i \succeq 0, ~ i=1,2,...,K$
\item $\sum_{i=1}^{K} \tQ_i \preceq {\bf S} $ and $\sum_{i=1}^{K} \hQ_i \preceq {\bf S} $.
\item $\mbox{Tr} \left ( \tL_i \tQ_i \right) = 0$ and $\mbox{Tr} \left ( \hL_i \hQ_i \right) = 0$.
\item $\mbox{Tr} \left ( \tD \left ( \sum_{i=1}^{K} \tQ_i - {\bf S} \right ) \right) = 0$.
\item $\mbox{Tr} \left ( \hD \left ( \sum_{i=1}^{K} \hQ_i - {\bf S} \right ) \right) = 0$.
\item $ r_i \nabla_i v_i(\tQ) + \tL_i - \tD = 0$
\item $ r_i \nabla_i v_i(\hQ) + \hL_i - \hD = 0$.
\end{enumerate}
Now multiplying (f) and (g) with $ (\hQ_i - \tQ_i) $ and $ (\tQ_i - \hQ_i) $ respectively, summing on $i$ and taking the trace we get
\begin{align}
0 & = \sum_{i=1}^{K} \mbox{Tr} \left [ (\hQ_i - \tQ_i)(  r_i \nabla_i v_i(\tQ) + \tL_i - \tD ) \right ] \nonumber \\
  & ~~~~ + \sum_{i=1}^{K} \mbox{Tr} \left [ (\tQ_i - \hQ_i)(  r_i \nabla_i v_i(\hQ) + \hL_i - \hD ) \right ] \nonumber \\
  & = \sum_{i=1}^{K} \mbox{Tr} \left [ (\hQ_i - \tQ_i) r_i \nabla_i v_i(\tQ) + (\tQ_i - \hQ_i) r_i \nabla_i v_i(\hQ) \right ] \nonumber \\
  & ~~~~ + \sum_{i=1}^{K} \mbox{Tr} \left [ (\hQ_i - \tQ_i) ( \tL_i - \tD )  +  (\tQ_i - \hQ_i)( \hL_i - \hD ) \right ] \nonumber \\
  & = \alpha + \beta
\end{align}
Re-arranging and evaluating the second term,
\begin{align}
& \beta = \textrm{Tr} \left [ \sum_{i=1}^{K} ( {\tilde Q}_i - {\hat Q}_i ) \left \{ (\tD - \tL_i) - (\hD - \hL_i)  \right \}  \right ] \nonumber \\
& ~ \stackrel{(c)}{=} \textrm{Tr} \left [ \sum_{i=1}^{K} ( \tQ_i \tD  - \tQ_i \hD + \tQ_i \hL_i - \hQ_i \tD + \hQ_i \tL_i + \hQ_i \hD ) \right ] \nonumber \\
& ~\stackrel{(d,e)}{=} \textrm{Tr} \left [ {\bf S} \tD + {\bf S} \hD \right] - \textrm{Tr} \left [ \sum_{i} \tQ_i \hD  +  \sum_{i} \hQ_i \tD  \right ] \nonumber \\
& ~ \quad \quad + \textrm{Tr} \left [ \sum_{i} (\tQ_i \hL_i + \hQ_i \tL_i)  \right ] \nonumber \\
& ~ \stackrel{(a)}{\geq} \textrm{Tr} \left [ \left ({\bf S} - \sum_{i} \tQ_i \right ) \hD \right ] + \textrm{Tr} \left [ \left ({\bf S} - \sum_{i} \hQ_i \right ) \tD \right ] \nonumber \\
& ~ \stackrel{(b)}{\geq} 0.
\end{align}
We have shown that $ \beta \geq 0 $ and hence for $ \alpha + \beta = 0 $ we need that $ \alpha \leq 0 $. Now
\begin{align}
\alpha & = \sum_{i=1}^{K} \mbox{Tr} \left [ (\hQ_i - \tQ_i) r_i \nabla_i v_i(\tQ) + (\tQ_i - \hQ_i) r_i \nabla_i v_i(\hQ) \right ] \nonumber \\
       & = \mbox{Tr} \left [ (\hQ - \tQ)^{T} g(\tQ,\underline{r}) + (\tQ_i - \hQ_i)g(\hQ,\underline{r}) \right ].
\end{align}
This condition is exactly the DSC condition defined in Section II. If for the broadcast channel game, we have that $ \alpha > 0 $ for some $\underline{r}$ then we have arrived at a contradiction and hence there cannot exist multiple NoE for that $\underline{r}$. For all such $\underline{r}$, $ \alpha > 0 $ is a sufficient condition for the uniqueness of the NoE.
\end{IEEEproof}

Note that we have derived the sufficient condition for uniqueness of the NoE for the broadcast channel game with joint covariance constraints. We state without proof that the DSC condition with matrix valued strategies holds for a more general scenarios (for example, sum power or per-antenna power constraint or other common constraints) and is not restricted to the broadcast channel problem (see \cite{Rosen65} for a discussion on how the DSC condition with vector valued strategies holds for concave games in general).

We now consider special cases for the broadcast channel and determine which values of $\underline{r}$ result in unique NoEs. We first state two trace inequalities that will be used to derive the uniqueness results.

\begin{lemma} \cite{Belmega10ineq}
For any positive integer $K$ and a set of positive semi-definite matrices $A_1,A_2,...,A_K$ and $B_1,B_2,...,B_K$ such that $A_1 \succ 0$ and $B_1 \succ 0$, we have that
\begin{align}
\textrm{Tr} \left \{ \sum_{k=1}^{K} (A_k - B_k) \left [ \left (\sum_{l=1}^{k} B_l  \right )^{-1} - \left (\sum_{l=1}^{k} A_l \right )^{-1}  \right ]  \right \} \geq 0.
\label{e:traceineq}
\end{align}
\end{lemma}
Note that the set of inequalities may not be the tightest trace inequalities. For example, for $K = 2$ and any positive real number $w$, it has been shown in \cite{Furuichi11ineq} that
\begin{align}
\mbox{Tr} &  [ (A_1 - B_1) ( B_1^{-1}  - A_1^{-1} )  \nonumber \\
     & + 4 (A_2 - B_2)\left \{ (wB_1 + B_2)^{-1} - (wA_1 + A_2)^{-1}  \right \} ] \geq 0.
\label{e:traceineq_2}
\end{align}
Clearly, there is much room for deriving generalizations of such inequalities and such generalizations will improve the characterization of the unique NoEs. We restrict our attention to the above known inequalities in this paper.

\subsection{Uniqueness results for normalized Nash equilibrium points}
\begin{prop}
For the aligned and degraded broadcast channel (ADBC) with dirty paper coding at the transmitter and interference canceling receivers, a unique normalized equilibrium point exists for $r_1 \geq r_2 \geq ... \geq r_K > 0$.
\label{lemma:unique_K_ADBC}
\end{prop}
\begin{IEEEproof}
For the ADBC, the utility obtained by the $k^{th}$ receiver is given by \cite{Weingarten06}
\begin{equation}
v_k(Q_k,Q_{-k}) = \textrm{log} \left ( \frac{| \sum_{i=1}^{k} Q_i + N_k |}{|  \sum_{i=1}^{k-1} Q_i + N_k |} \right ).
\end{equation}
For this utility function we now show that $ \alpha > 0 $ and thus determine the unique NoEs. Let $(\tQ_1,\tQ_2,...,\tQ_k)$ and $(\hQ_1,\hQ_2,...,\hQ_k)$ be any two tuples of covariance matrices which satisfy the covariance constraint:
$ \sum_{i=1}^{K} \tQ_i \preceq {\bf S} $ and $ \sum_{i=1}^{K} \hQ_i \preceq {\bf S} $. Substituting the utility obtained when using the two sets of covariances in the DSC condition, we get
\begin{align}
& \textrm{Tr} \left [ \sum_{k=1}^{K} r_k ( {\hQ}_k - {\tQ}_k ) \left \{  \nabla_k v_k({\tQ}) - \nabla_k v_k({\hQ})   \right \} \right ] 
\end{align}
\begin{align}
& = \textrm{Tr} \Bigg [ \sum_{k=1}^{K} r_k ( {\hQ}_k - {\tQ}_k ) \bigg \{ (N_k + \sum_{i=1}^{k} \tQ_i  )^{-1}  \nonumber \\
& ~~~~~~~~~~~~~~~~~~~~~~~~~~~~~~~~~~~~~~~~ - (N_k + \sum_{i=1}^{k} \hQ_i  )^{-1} \bigg \} \Bigg ] \nonumber \\
& = \sum_{n=1}^{K-1} (r_{n} - r_{n+1}) \mT_n + r_K \mT_{K},
\end{align}
where the term $\mT_n$ can be expressed as
\begin{equation}
\textrm{Tr} \Bigg [  \sum_{k=1}^{n} (\hQ_k - \tQ_k) \bigg \{ (N_k + \sum_{i=1}^{k} \tQ_i)^{-1} - (N_k + \sum_{i=1}^{k} \hQ_i)^{-1} \bigg \} \Bigg ].
\end{equation}
It is now sufficient to show that $\mT_n > 0 $. Notice that the structure of $\mT_n$ closely resembles the inequality in (\ref{e:traceineq}). Choose the quantities $A_1 = N_1 + \tQ_1$, $B_1 = N_1 + \hQ_1$, $A_i = N_i - N_{i-1} + \tQ_i$ and $B_i = N_i - N_{i-1} + \hQ_i$. By definition, since $N_1$ is a positive definite matrix and $\tQ_1$, $\hQ_1$ are positive semi-definite the matrices $A_1$ and $B_1$ are strictly positive definite. From the degradedness of the channel, we get that $N_i - N_{i-1}$ is a positive semi-definite matrix and hence $A_i$ and $B_i$ are positive semi-definite for $i=2,...,K$. Substituting the values of $A_i$ and $B_i$ in Eq. \ref{e:traceineq}, it is straight forward to see that $\mT_n \geq 0$. For an ADBC channel having identity channel matrices, we know from \cite{Belmega10} that if the NoEs $\tQ \neq \hQ$ then $ \mT_n > 0 $ and hence the normalized equilibrium points of the ADBC game as unique for $r_1 \geq r_2 ... \geq r_K > 0 $.
\end{IEEEproof}

It is clear that the region of weight vectors $r$ for which uniqueness can be shown is dependent on the tightness of the matrix trace inequalities. For $K=2$, the inequality in (\ref{e:traceineq}) has been improved to the inequality in (\ref{e:traceineq_2}). Thus for the 2-user ADBC, the uniqueness can be derived for a more general set of weight vectors.

\begin{lemma}
For a 2-user ADBC with dirty paper coding at the transmitter and interference canceling receivers, a unique normalized equilibrium point exists for $r_1 \geq r_2/4 > 0$.
\end{lemma}
\begin{IEEEproof}
The proof follows exactly on the lines of Lemma \ref{lemma:unique_K_ADBC} with the DSC condition decomposing into two terms given by $(r_1 - \frac{r_2}{4}) T_1$ and $\frac{r_2}{4} T_2$. Now using the inequality in (\ref{e:traceineq_2}) with $w=1$, it is easy to show that there exists a unique normalized equilibrium point for each point in the region $r_1 \geq r_2/4 > 0$.
\end{IEEEproof}

\subsection{Discussion}
We have characterized the unique NoEs of the ADBC above. Uniqueness of the NoEs ensures that each there exists a single reasonable outcome for the broadcast channel game and simplifies the development of algorithms to compute NoEs. We also note that the proof of uniqueness makes explicit use of the degradedness of the broadcast channel and hence cannot be directly extended to the AMBC and the general GBC. Characterizing the uniqueness of the NoEs for the AMBC and the GBC and computation of these equilibria \cite{Pang08intfchannel} is currently being investigated .

\section{Conclusions}
\label{sec:conclusions}
Strategic behavior among rational non-cooperative receivers in a Gaussian broadcast channel has not been studied so far in literature. In this paper, we presented a game theoretic model for the general Gaussian broadcast channel and showed that it belongs to the special class of generalized Nash equilibrium problems with common dependent constraints. We showed the existence of the normalized equilibrium points, the solutions of the GNEP with coupled constraints, for the Gaussian broadcast channel with dirty paper coding and linear precoding strategies. In general, there exist multiple equilibrium points for a GNEP. We then derived a sufficient condition to determine the unique normalized equilibrium points and characterized the uniqueness of these points for the special case of an aligned and degraded broadcast channel.  

%\IEEEtriggeratref{2}

\bibliographystyle{IEEEtran}
\bibliography{references}

\end{document}